\documentclass[aps,prl,reprint]{revtex4-1}
\usepackage{blindtext}
\usepackage{anysize}
\usepackage{layout}
\usepackage{latexsym}
\usepackage{epsfig}
\usepackage{amssymb}
\usepackage{amsmath}
\usepackage{color}
\usepackage{graphicx}
\usepackage{setspace}
\usepackage[font={small},format=plain]{caption}
 \usepackage[T1]{fontenc}
 \usepackage{helvet}
\usepackage[margin=1.1in]{geometry}

\newcommand{\beq}{\begin{equation}}
\newcommand{\eeq}{\end{equation}}
\newcommand{\be}{\begin{eqnarray}}
\newcommand{\ee}{\end{eqnarray}}

\newcommand {\bea}{\begin{eqnarray}}
\newcommand {\eea}{\end{eqnarray}}

\begin{document}
\title{Comment on R. Golestanian's ``Enhanced Diffusion of Enzymes that Catalyze Exothermic Reactions"}
\author{K. Tsekouras$^{1}$, C.Riedel$^{2}$, R.Gabizon$^{2}$, S.Marqusee$^{2,3}$, S.Press\'e$^{1,4\ast}$
\\
and C.Bustamante$^{2,3,5,6,7,8\ast}$}
\begin{abstract}
{\small{
\hspace{-0.125 in}$^{1}$Department of Physics, IUPUI Indianapolis IN 46202 USA
\\
$^{2}$California Institute for Quantitative Biosciences, QB3, UC Berkeley, CA 94720 USA  
\\
$^{3}$Department of Molecular and Cell Biology, UC Berkeley, CA 94720, USA 
\\
$^{4}$Department of Cellular \& Integrative Physiology, IU School of Medicine, Indianapolis IN 46202 USA 
\\
$^{5}$Jason L. Choy Laboratory of Single-Molecule Biophysics and Department of Physics, UC Berkeley, CA 94720, USA
\\
$^{6}$Department of Chemistry, UC Berkeley, CA 94720, USA
\\
$^{7}$Howard Hughes Medical Institute, UC Berkeley, CA 94720, USA
\\
$^{8}$Kavli Energy Nano Sciences Institute, UC Berkeley and Lawrence Berkeley National Laboratory, CA 94720, USA.}
\\
$\ast$corresponding author}
\end{abstract}
\maketitle

We \cite{riedel_heat_2015} as well as others \cite{sengupta_enzyme_2013, muddana_substrate_2010,dey_chemotactic_2014} have
shown that some active enzymes exhibit enhanced diffusion. 
In the process of building a model for this effect,  we eliminated alternative explanations through experimental controls; see  Ref.~\cite{riedel_heat_2015} and SI. 

In a recent (2015) PRL by R. Golestanian \cite{golestanian_enhanced_2015}, the author theoretically examines multiple 
explanations for increased diffusion of enzymes that catalyze highly exothermic reactions.  
One of these is \textit{collective heating} (CH) that
attributes the active enzyme's enhanced diffusion to a temperature rise of the buffer caused by the accumulation of reaction heat in the center of the reaction container. 
Golestanian concludes that CH, 
is the best candidate explanation for our results recently published in Nature \cite{riedel_heat_2015}.
Here we present evidence to counter this claim.

Our controls rule out  the possibility that global or local heating of the solvent surrounding the enzyme in our experiments is responsible for enhanced enzyme diffusion, as predicted from a heat capacity calculation \cite{riedel_heat_2015}.

Briefly, assuming perfect thermal isolation, we calculate the expected maximum temperature rise, $\Delta T$, anywhere within the container, assuming total substrate depletion 
for our most exothermic enzyme, catalase. Importantly, substrate is \textit{not} replenished in our experiments. Using $[C(H_2O_2)]=25mM$ as the concentration of substrate (which for catalase is hydrogen peroxide), $C_p=4.18\cdot 10^3 J/(KL)$ as the water heat capacity and $\Delta H=100 kJ/mol$ as the enthalpy of the reaction, we get:
\beq
\Delta T = [C({H_2O_2})]\Delta H/C_p\approx 0.6K. 
\eeq
Interpolating the 300.6K viscosity from experimental values \cite{A_john_dre} 
and using the Stokes-Einstein's relation yields this ratio
of enhanced ($T=300.6K$) to unenhanced ($T=300K$) diffusion coefficient:
\beq
\frac{D'}{D} = \frac{kT'}{6\pi\eta' r}\frac{6\pi\eta r}{kT}\rightarrow D' = D\frac{T'\eta}{T\eta'}\approx 1.015D.
\label{drif}
\eeq
Note that $300.6K$ is \textit{not} the spatially averaged temperature rise within the container. Instead it is the maximum temperature rise possible \textit{at any point} inside the experimental container irrespective of the speed with which the reaction happens. Since substrate is depleted and not replenished, no more heat can ever be produced within the container.
What is more, this temperature rise is totally insufficient to explain the $25\%$ diffusion coefficient rise for catalase.

If, instead, we choose to calculate the temperature rise expected to occur during our experiment, we employ catalase's concentration ($[C_{cat}]=1nM$) and catalytic rate ($k=2.5\cdot 10^4 s^{-1}$ that we measured at $300K$ at saturation) plus the experiment duration ($t=30s$), and the number of catalase's catalytic sites ($n=4$), to find: 
\beq
\Delta T= \frac{\Delta Q}{C_p} = \frac{nk t[C_{cat}]\Delta H}{C_p}\approx 0.072K
\label{dT}
\eeq
where once more we assumed no heat loss. Interpolating the 300.072K viscosity from experimental values \cite{A_john_dre} as before
and using again Eq.\ref{drif} yields this ratio of enhanced ($T=300.072K$) to unenhanced ($T=300K$) diffusion coefficient: 
$D' = D\frac{T'\eta}{T\eta'}\approx 1.00112D$. If we take the Arrhenius rate enhancement into consideration, we get $D'\approx 1.00118D$.
Diffusion coefficient increases of $0.0112\%$ or $0.00118\%$ are also nowhere near the $25\%$ observed for active labeled catalase. Our measurements were taken during a window of just $30 s$, starting $10 s$ after the enzyme was added to the solvent-substrate mix. In all cases, unavoidable heat loss further reduces these small temperature changes.  

In order to obtain a larger diffusion coefficient enhancement, 
Golestanian ignores substrate depletion and assumes steady state (and thus replenishment of substrate). 
In our experiments, since we never replenish substrate, the only steady state that can be reached is when substrate has been depleted. 

We add finally that CH cannot explain why labeled active urease shows a larger diffusion increase than  
labeled inactive urease placed in the same container as unlabeled active catalase, despite active catalase producing more heat than active urease
as seen in Fig. 3 of the Ref.~ \cite{riedel_heat_2015} SI. 
Nor can CH explain why labeled inactive urease in the same container as unlabeled active catalase shows smaller diffusion rises than labeled active catalase, despite the fact that in identical containers the same amounts of active catalase produce identical amounts of heat. 

In addition, a simple (Gaussian) likelihood ratio test 
reveals that our control (Ext. Data Fig.3 \cite{riedel_heat_2015}) is 22$\times$ more likely to show no trend 
(fixed urease diffusion at $31.5 \mu m^{2}/s$)
than a $25\%$ increase
(as would be expected in the presence of active catalase had Golestanian's CH hypothesis held for our experiment).

Finally, ignoring our own experiments altogether, Dey \textit{et al.} \cite{dey_chemotactic_2014} show enzymes in the same device separating out in physical space on the basis of  catalytic activity.  If CH held for these experiments, and since all enzymes share the same container, all enzymes would experience the same heating and exhibit size, not activity, dependent diffusion; so CH cannot apply to those experiments either.

{\bf Summary:}
Collective heating cannot explain our \cite{riedel_heat_2015} and other \cite{dey_chemotactic_2014} experiments on enhanced enzyme diffusion. 
This said, collective heating could very well apply in systems where many orders of magnitude more heat is produced.

{\bf Acknowledgements}
We thank A. Sen, G. King, D. Makarov, A. Lee and K. Ghosh for helpful discussions.
We acknowledge support from NIH grants R01-GM0325543 (C.B.) and R01-GM05945 (S.M.), the US DoE, Office of Basic Energy Sciences, Division of Materials Sciences and Engineering under contract no. DE-AC02-05CH11231 (C.B.), the NSF grants MCB-1412259 (S.P.) and MCB-1122225 (S.M.), the Human Frontier Science Program (C.R) and the Burroughs-Wellcome Fund (S.P).

\bibliographystyle{apsrev4-1}
\bibliography{furca}

\singlespacing
\thispagestyle{plain}
\pagestyle{plain}


\end{document}